\documentclass[preprint]{emulateapj} 
\begin{document}

\title{ Dynamical measurements of the interior structure of exoplanets}  

\author{Juliette C. Becker$^1$ \& Konstantin Batygin$^2$} 

\affil{$^1$Cahill Center for Astronomy and Astrophysics, California Institute of Technology, 1200 E. California Blvd., Pasadena, CA 91125} 
\affil{$^2$Institute for Theory and Computation, Harvard-Smithsonian Center for Astrophysics, 60 Garden St., Cambridge, MA}
\email{jbecker@caltech.edu}

\begin{abstract} 

Giant gaseous planets often reside on orbits in sufficient proximity to their host stars for the planetary quadrupole gravitational field to become non-negligible. In presence of an additional planetary companion, a precise characterization of the system's orbital state can yield meaningful constraints on the transiting planet's interior structure. However, such methods can require a very specific type of system. This paper explores the dynamic range of applicability of these methods and shows that interior structure calculations are possible for a wide array of orbital architectures. The HAT-P-13 system is used as a case study, and the implications of perturbations arising from a third distant companion on the feasibility of an interior calculation are discussed. We find that the method discussed here is likely to be useful in studying other planetary systems, allowing the possibility of an expanded survey of the interiors of exoplanets.  
\end{abstract} 

\maketitle

\section{Introduction}

Understanding the physical structure of giant exoplanets holds great scientific value. In addition to being a subject of considerable interest in itself, such knowledge can shed light on topics such as the behavior of materials under high pressures \citep{1982AREPS..10..257S, 1999Sci...296...72G}, the dominant process responsible for planet formation \citep{1996Icar..124...62P}, as well as the mechanism behind hot Jupiter radius inflation \citep{2001ApJ...548..466B, 2002A&A...385..166S, 2007ApJ...668L.171B,2010ApJ...714L.238B}. Simultaneously, studying distant planets is made difficult by the scarcity of direct methods: while planets in our own solar system are close enough for probes to be sent for the purpose of studying their properties \citep{2011Icar..216..440H}, this is not feasible for exoplanets. Instead, we must rely on indirect methods to obtain observationally elusive information. The focus of this paper is the determination of both the robustness of such methods and the dynamical range of their applicability. 

In planar two-planet systems, tidal forces can cause planetary orbits to attain orbital equilibration \citep{2002ApJ...564.1024W, 2007MNRAS.382.1768M}. Specifically, while conserving the total angular momentum, tidal dissipation results in the decay of the orbital eccentricities and semi-major axes \citep{1963MNRAS.126..257G, 1980A&A....92..167H}.  Provided a sufficient amount of time (on the order of a few circularization timescales), the interplay between secular planet-planet interactions and tidal damping results in a stationary orbital state characterized by apsidal alignment (and co-precession) of the orbits. Because of apsidal alignment, the associated variations in the eccentricities also vanish \citep{1999ssd..book.....M}. As a result, such an orbital state is often referred to as the "fixed point" \citep{2007MNRAS.382.1768M}. 

The rate of apsidal precession\footnote{Throughout this paper, we use the terms apsidal motion and precession of the perihelion interchangeably.} of a close-in planet is in part determined by the planetary Love number, $k_{2}$, which measures the planetary quadrupole potential that arises from tidal deformation \citep{1939MNRAS..99..670S}. In fact, for hot Jupiters, tidal precession tends to dominate over other non-Newtonian effects such as General Relativity \citep{2009ApJ...698.1778R}. Since the Love number quantifies a planetÕs susceptibility to elongation due to tidal forces, it is inherently related to its interior mass distribution. This means that in a multi-planet system, the inner-most planet's degree of central concentration determines the quantitative nature of the fixed point onto which the system eventually settles. Thus, there is a clear, direct relationship between the interior structure and the orbital state. More specifically, the observable quantities of transiting extrasolar planets (e.g. eccentricity, mass, semi-major axis, radius) can be used to determine the Love number $k_{2}$, by requiring fixed point conditions (i.e. apsidal alignment and co-precession) to be satisfied. In turn, the Love number, although an inherently degenerate quantity \citep{2011A&A...528A..18K}, can be used to constrain the interior mass distribution of the planet with the aid of thermal evolution calculations, by requiring that the interior models yield the correct degree of central concentration \citep{2009ApJ...704L..49B, 2012A&A...538A.146K}. 

The method for interior structure determination described above is subject to a number of assumptions. First, the planetary pair in question is assumed to be co-planar. Mardling (2010) showed that if the inclination between the planets is significant, the orbits never settle onto a fixed point, and instead converge onto a limit cycle characterized by periodic oscillations in eccentricity and inclination. Furthermore, such a limit cycle is generally long-lived (the mutual inclination decays slowly compared to the circularization timescale) rendering the interior structure calculation inapplicable in mutually inclined systems.  Second, even if the system is planar but comprises more than two planets, orbital equilibration may require an unreasonably long time \citep{2011ApJ...730...95B}, again preventing planetary orbits from settling onto a fixed point. At the same time, if the inner-most planetary pair is sufficiently isolated (i.e. the timescale of external perturbation greatly exceeds that for the self-interaction of the planetary pair), adiabatic theory (Henrard 1993)  suggests that the deviations away from the fixed point should be negligible. As already mentioned above, delineating the requirements for adiabatic behavior of fixed points is the primary aim of this paper. 

Upon a cursory inspection, it appears that planets HAT-P-13b and HAT-P-13c \citep{2009ApJ...707..446B} reside at a fixed point, and to date, HAT-P-13 remains the only system to which the above-described analysis has been applied. Interestingly, follow up radial velocity observations have revealed evidence of the existence of a third massive distant planet, HAT-P-13d \citep{2010ApJ...723L.223W}, casting doubt on the stationary nature of the inner planet pair. Particularly, it is no-longer clear if apsidal alignment of planets b and c is truly a result of tidal evolution or if the system is being observed at an aligned phase of a circulating/librating cycle. Due to this uncertainty, the HAT-P-13 system presents an illustrative example, to which the theoretical arguments developed here can be applied. 

The paper is structured as follows. In section 2, we discuss the degeneracy among various planetary interior models and demonstrate the utility of the Love number as a means of differentiating among them. In section 3, we consider the dissipative evolution of a hot Jupiter under secular perturbation from a precessing massive companion and demonstrate the system's convergence onto a fixed point. In section 4, we consider the dissipative evolution of a hot Jupiter under perturbations from a system of two interacting planets and derive conditions, under which the system settles onto an adiabatic fixed point. In section 5, we apply the developed theory to the HAT-P-13 system and consider the stability of the 3-planet configuration. We summarize and discuss our results in section 6.

\section{Constraints on Extrasolar Planetary Structures}

Observational constraints on the mass and radius of a transiting planet do not determine the interior structure of the planet. In fact, assuming no significant interior heat sources\footnote{As discussed by \cite{2010SSRv..152..423F} and the references therein, a significant fraction of hot Jupiters require an interior heat source to maintain inflated radii. Still, it is useful to consider an evolved, degenerate planet as a guiding example.}, the radii of evolved (i.e. degenerate) gas giant planets are largely independent of their masses \citep{1982AREPS..10..257S}. Instead, the radius of a gas giant is primarily dictated by its composition, or what mass-fraction of heavy elements it contains. The decrease in radius, associated with an increased proportion of heavy elements is roughly the same whether the heavy elements form a core or are spread throughout the gaseous envelope. Accordingly, an enhanced Helium fraction has been used as a means of mimicking the presence of a core in thermal evolution models (see for example Burrows et al 2007). This point is of considerable importance, since \cite{2012ApJ...745...54W} suggest that MgO and H$_{2}$O are both soluble in metallic hydrogen at high pressures and temperatures (as would be found in the core of such a gas giant), suggesting in turn that the cores of giant planets could be evaporating.
Although two planets, one with and one without a core, could have the same radius, their interior structures clearly differ. Therefore, an additional constraint (the Love number) is needed to differentiate between the two envisioned interior states.  

Examples of state-of-the-art numerical interior models where the Love number is used as a constraint already exist in the literature \citep{2009ApJ...704L..49B, 2012A&A...538A.146K}. Here, we shall revisit the calculation with a closed-form analytical interior model.

To illustrate that a measurement of mass and radius alone is not sufficient to determine the interior structure of a planet, we will develop two valid models of planetary structure with the same radius but different interior structures. To a fair approximation, the equation of state of a degenerate giant planet can be represented by the $n=1$ polytrope \citep{1974AZh....51.1052H}: 
\begin{equation}
P = K (1-y^2) \rho^2 
\end{equation}
where $P$ is pressure, $\rho$ is density, $y$ is the heavy-element mass-fraction and $K$ is a constant. For this equation of state, the hydrostatic equation:
\begin{equation}
\frac{dm}{dr} = \frac{-1}{G} \frac{d}{dr} ( \frac{r^{2}}{\rho} \frac{dP}{dr})  
\end{equation}
yields a closed-form solution for the density profile:
\begin{equation}
\rho = A   \frac{\sin (kr)}{kr} + B  \frac{\cos(kr)}{kr}
\label{dens_rela}
\end{equation}
where $k = \sqrt{\frac{2 \pi G}{K}}$. If the planet in question has no core, the $r = 0$ boundary condition requires $B=0$, and $A=\rho_{c}$, the central density of the planet. This yields a density profile for a planet without a core:
\begin{equation}
\label{coreless}
\rho_{\rm{coreless}}(r) =  \rho_{\rm{c}} \frac{\sin (kr)}{kr}  
\end{equation}

From that density profile, we can write the radius of the coreless planet:
\begin{equation}
R_{\rm{coreless}}=\sqrt{\frac{\pi \ K (1-y)^{2}}{2 G}}.
\end{equation}
Note that the above expression is independent of the planetary mass, $m$ \citep{1969ApJ...158..809Z}.

Let us now turn our attention to a planet of identical mass and radius, but with $y = 0$ and a heavy-element core with a constant density. To derive the density profile for the cored case, we consider a planet with identical mass and radius to the coreless planet. In the cored case, the definition of hydrostatic equilibrium can be used to show that $B \simeq {m_{\rm{core}} k^{3}}/4\pi$. Similarly, the value of A for the cored case (which we will denote $A_{core}$) can be found by solving the following integral for the mass of a cored planet in the envelope. 
\begin{eqnarray}
\label{tofindA}
\int_{R_{\rm{core}}}^{R} (A_{\rm{core}} \frac{\sin (k R )}{k R} + \frac{m_{\rm{core}} k^{3}}{4\pi}  \frac{\cos(k R)}{k R}) r^{2} dr \nonumber \\ = \frac{M - M_{core}}{4 \pi}
\end{eqnarray}
The value of $A_{\rm{core}}$ can be used to find the value of $m_{\rm{core}}$ by matching the values of $R$ and $m$ for the cored and coreless cases, thereby satisfying the boundary condition:
\begin{equation}
A_{\rm{core}} \frac{\sin (k R )}{k R} = - \frac{m_{\rm{core}} k^{3}}{4\pi}  \frac{\cos(k R)}{k R}
\end{equation}
The value of $A_{\rm{core}}$,  together with the value of $B$ found from hydrostatic equilibrium, yield the complete expression for the density profile of a planet with a core:
\begin{equation}
\label{core}
\rho_{\rm{cored}}(r) = \\
\left\{  \begin{array}{c} 
\rho_{\rm{core}}\ \ \ \ \ \   \rm{if}\ r \leq R_{core} \\
A_{\rm{core}} \frac{\sin (kr)}{kr} + \frac{m_{\rm{core}} k^{3}}{4\pi}  \frac{\cos(kr)}{kr}
           \end{array}  \right.  \ \ \  \rm{if}\ r > R_{\rm{core}} 
\end{equation}

Equations \ref{coreless} and \ref{core} show the difference in density profile between two planets identical in mass and radius, between which the only difference is whether the heavy-elements manifest in a core or are spread throughout the envelope. 

Sterne (1939) used tesseral harmonic functions to write down the following differential equation for interior characterization. 
\begin{eqnarray}
\label{etaeqn}
\int_0^{R_{\rm{planet}}} (r \eta'(r) + \eta(r)^{2} - \eta(r) -6 \nonumber \\
+ 6 \frac{\rho(r)}{\rho_{m}(r)}(\eta(r) +1)) \,  \mathrm{d}\eta(r) =0
\end{eqnarray}
This differential equation relates the density distribution of a planet to the dimensionless quantity $\eta$, where $\rho_{m}$ is the average density interior to a given shell at $r$. The solution for $\eta$ can in turn be used to find the Love number. 
\begin{equation}
\label{Loveno}
k_{2}=\frac{3-\eta(R_{planet})}{2+\eta(R_{planet})}
\end{equation}   
Note that the Love number ($k_{2}$) is dependent on the density profile, as the Love number is found from the quantity $\eta(r)$ given by equation (\ref{etaeqn}). 

Different density profiles, then, correspond to different Love numbers, even for an identical pairing of mass and radius. This is illustrated by the two cases considered in this section. Regardless of the heavy element content, the density profile for the coreless case gives the well-known value of $k_2 = 0.52$ \citep{1959cbs..book.....K, 2002ApJ...564.1024W, 2009ApJ...698.1778R}. In contrast, the Love number of the cored case varies with the mass of the planet's core. 

\begin{figure}[htbp] 
   \centering
   \includegraphics[width=3.4in]{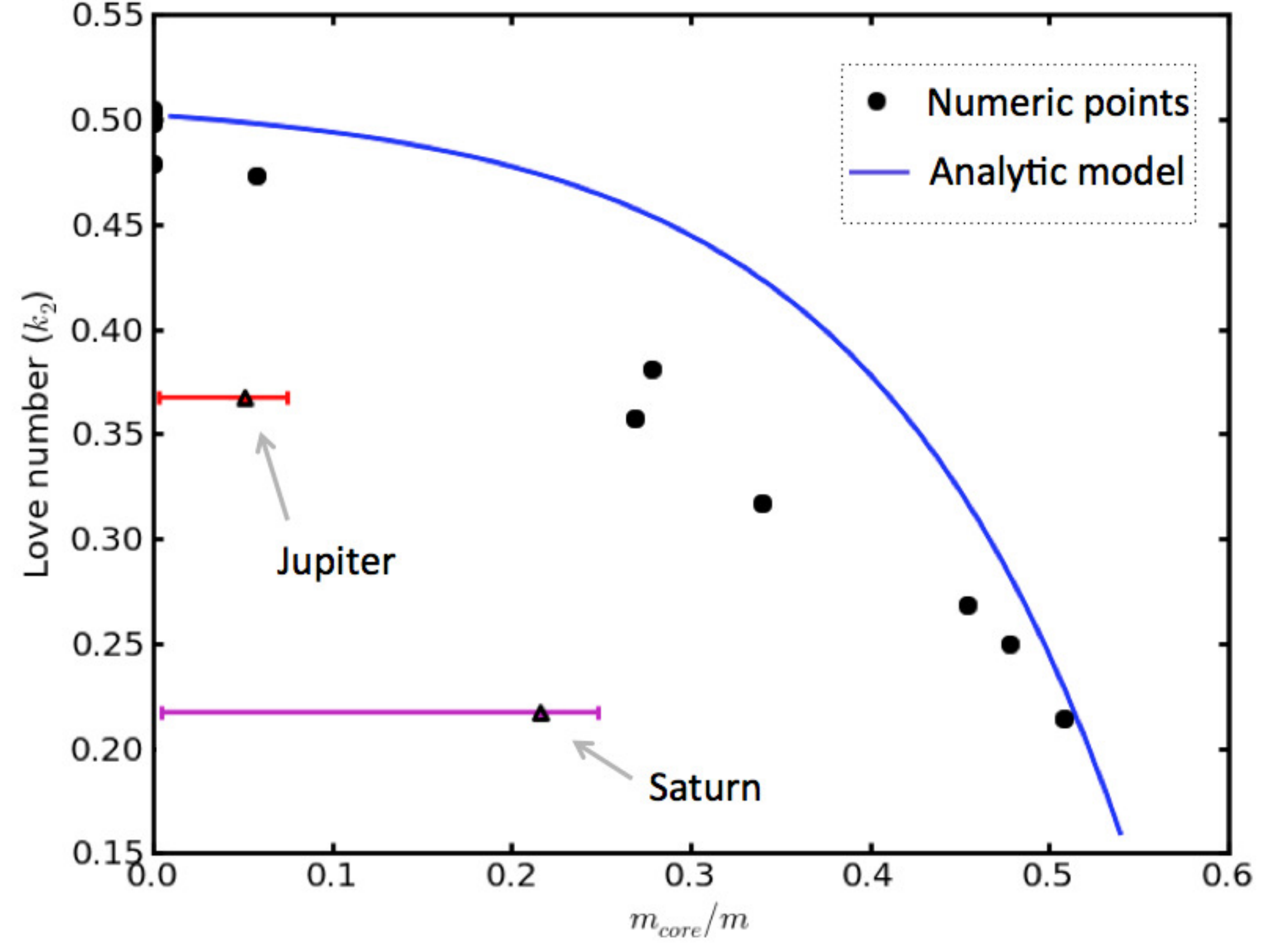} 
   \caption{$k_{2}$ as a function of $m_{\rm{core}} / m$: black points are the results of numerical thermal evolution models, while the blue curve is the analytical model described here. The agreement between the analytical model and the numerical calculations is satisfactory. Also plotted for comparison are the approximate Love numbers of Jupiter and Saturn \citep{2005GeCoA..69.4173R,1999P&SS...47.1183G}}   \label{k2plot}
\end{figure}

As demonstrated in Figure \ref{k2plot}, the planetary Love number, $k_2$ can be expressed as a function of the ratio $R_{\rm{core}}/R$ (which, for a given core density, is equivalent to the quantity $m_{\rm{core}}/m$). Specifically, the polytropic model formulated above, proceeding with the distribution defined in equation (\ref{core}), is plotted as a blue curve.

The same comparison can be made within the context of a more sophisticated interior model. Using the MESA stellar and planetary evolution code \citep{2010ascl.soft10083P}, we have compiled a suite of evolved (i.e. age $= 5$ Gyr) interior models for planets with Jupiter's mass and radius but differing Helium fractions (ranging $Y = 0.2-0.4$), core masses (ranging $m_{\rm{core}} = 0 - 0.5 m_{\rm{Jup}} $) and interior heating rates (ranging $\dot{\epsilon} = 0 - 4 \times 10^{25}$ ergs/s). The specific input parameters of the interior models as well as their corresponding Love numbers are listed in Table \ref{tableint} and are plotted as points in Figure \ref{k2plot}. For all numerical calculations, we used the the tabulated SVHC Hydrogen-Helium equation of state (Saumon 1995) and a constant density core ($\rho_{\rm{core}} = 3$ g/cc) where all of the imposed interior heating was concentrated. For the atmospheric boundary condition, we used the analytic semi-gray model formulated by \citep{2011A&A...527A..20G}, choosing the same opacities used by \cite{2010A&A...520A..27G}  to match the numerical atmospheric models of \cite{2008ApJ...678.1419F} and setting the irradiation temperature to $T_{\rm{irr}} = 1500$K.

\begin{table}
\begin{center}
\begin{tabular}{l l l l l } 
 $Y$ & $M_{\rm{core}}$ ($M_{\rm{Jup}}$)  & $\dot{\epsilon}$ ergs/s  & $k_2$  \\
\hline
 $0.2$ & $0 $  & $0$ & $0.505$  \\
 $0.2$ & $ 0.194 $  & $0$ & $0.318$  \\
 $0.2$ & $0.430 $  & $0$ & $0.250$  \\
   $0.25$ & $0 $  & $0$ & $0.503$  \\
  $0.25$ & $0.105 $  & $0$ & $0.381$  \\
 $0.25$ & $0.105 $  & $1.5 \cdot 10^{25}$ & $0.358$  \\
  $0.25$ & $0.262 $  & $0$ & $0.268$  \\
 $0.25$ & $0.524 $  & $4.0 \cdot 10^{25}$ & $0.214$  \\
 $0.3$ & $0 $  & $0$ & $0.498$  \\
 $0.3$ & $0.001 $  & $1.5 \cdot 10^{25}$ & $0.473$  \\
 $0.4$ & $0 $  & $0$ & $0.479$  \\
 \tableline
\end{tabular}
\end{center}
\caption{Numerical Interior Models: With an additional level of degeneracy, these numerical results show overall agreement with the simpler polytropic model. } 
\label{tableint} 
\end{table}

While the numerical models highlight the effect of added degeneracy (that is, the interior heating), on the $k_2 - m_{\rm{core}}/m$ relationship, they also demonstrate the overall qualitative agreement between the state-of-the-art evolutionary calculations and the polytropic model considered above. Furthermore, it should be noted that the spread in the numerical points is not particularly severe implying that meaningful constraints on $m_{\rm{core}}/m$ can still be gleaned from $k_2$ even without proper knowledge of exact characteristic parameters.

As mentioned above, there exists a clear connection between the interior structure of the planet and its orbital state. Namely, tidal deformation of a planet gives rise to a quadrupole gravitational field that results in apsidal precession of the orbit. The tidal precession rate is directly proportional to the Love number, and to lowest order in eccentricity, is given by:
\begin{equation}
\label{tidalprec}
\left(\frac{d\varpi}{dt}\right)_{\rm{tidal}} = \frac{15}{2} n k_{2} \left(\frac{R}{a}\right)^{5} \frac{M}{m},
\end{equation} 
where $n$ is the mean motion, $a$ is the semi-major axis, and $M$ is the mass of the host star.

The relationship between the apsidal precession and the Love number, renders the direct measurements of the tidal deformation (such as that proposed by \cite{2011A&A...528A..41L}) unnecessary (although highly complementary) to the determination of planetary interior structure. Indeed, if the apsidal precession of an orbit can be constrained, then so too can the interior structure of the planet. This proportionality is the foundation of the methods described in the following section. 

\section{Dissipative Evolution of a Giant Planet, Perturbed by a Distant, Massive Object}

Prior to describing the mathematical formulation of the quantitative nature of the fixed point and its relationship to the planetary Love number $k_2$, it is first useful to consider the general category of orbital architectures where persistent apsidal alignment may be of interest. Because the perihelion cannot be defined for a circular orbit, the presence of eccentricity is desirable for an observational determination of its longitude. Simultaneously, significant tidal dissipation (that acts to circularize the orbits) is necessary for orbital equilibration. 

As discussed by \cite{2011ApJ...730...95B}, these contradictory factors imply that orbital architectures where apsidal alignment in presence of non-negligible eccentricity may be observed, are those where the inner orbit constitutes a relatively small fraction of the overall angular momentum deficit. Put simply, the eccentricity of a close-in orbit can only be maintained in face of tidal dissipation by a distant, eccentric, massive perturber. As a result, following \cite{2011ApJ...739...31L}, here we shall focus on characterizing the interactions among distant orbits and as a guiding approximation\footnote{Note that this is not a requirement of the considered theory. Rather, this is a matter of convenience aimed primarily at clarifying the mathematical representation of the orbital evolution.}.

In principle, the orbits we wish to focus on here are those not locked in low-order mean motion resonances i.e. period ratios that satisfy $P_b/P_c \simeq (u - v)/u$ where $(u,v) \in \mathbb{Z}, \  v \leqslant 2$. However, for simplicity, we shall go one step further and assume that the planetary semi-major axis ratio is small: $a_b/a_c \ll 1$. As was first shown by \cite{laplace} the calculation of the dynamical evolution of such orbits need not resolve the Keplerian motion of the planet because the deviation from Keplerian motion is only significant in presence of resonances. Instead, as argued from a more intuitive point of view by Gauss (1866) it is sufficient to treat the orbits as massive wires where the planetary mass is smeared out along the orbit with the line density inversely proportional to the orbital speed, and compute the secular gravitational interactions between them.  

The equilibration of a secular system due to tidal dissipation has already been discussed by a handful of authors, using different variants of perturbative secular theory and numerical methods \citep{2002ApJ...564.1024W, 2007MNRAS.382.1768M, 2010ApJ...708.1366V, 2011ApJ...730...95B, 2012A&A...538A.105L, 2012CeMDA.113..215V}. Here, we shall not attempt to improve the accuracy of the perturbative approach, but instead work in the opposite direction with the aim to write down a tractable set of equations while still capturing the essential features of the dynamical evolution. Indeed at an age when numerical \textit{N}-body software \citep{1998AJ....116.2067D,1999MNRAS.304..793C} and computational resources are readily available, this would seem to be the most fruitful approach to perturbation theory of this kind. 

To leading order in orbital eccentricities, assuming small inclinations, the Hamiltonian governing the secular motion of the planet $b$, perturbed by a distant body $c$, is given by \cite{1999ssd..book.....M} :
\begin{equation}
\label{Hsecb}
\mathcal{H}^{\rm{sec}}_b = \frac{A}{2} e_b^2 + A' e_b e_c \cos( \varpi_b - \varpi_c),
\end{equation}
where $e$ is the eccentricity and $\varpi$ is the longitude of perihelion. This formulation is generally referred to as the Laplace-Lagrance secular theory. The constants $A$ and $A'$ are exclusively functions of the planetary masses and the semi-major axes (which become constants of motion after averaging over the mean motion i.e. spreading the planetary mass across the orbit). For systems of interest to us, these interaction coefficients are given by:
\begin{eqnarray}
A &=& \left(\frac{d\varpi}{dt}\right)_{tidal} + \left(\frac{d\varpi}{dt}\right)_{GR} + \frac{3 }{4} \frac{m_{c}}{M} \left(\frac{a_b}{a_c} \right)^3 n_{b}  \nonumber \\
A' &=&-\frac{15}{16}  \frac{m_{c}}{M} \left(\frac{a_b}{a_c} \right)^4 n_{b}.
\end{eqnarray}
In the above expressions, we have taken advantage of the assumption of well separated orbits i.e. $a_b/a_c \ll 1$ and expanded the Laplace coefficients as hypergeometric series to leading order. Note that the tidal precession $d\varpi/dt_{\rm{tidal}}$, given by equation (\ref{tidalprec}) as well as the general relativistic precession $d\varpi/dt_{\rm{GR}} = 3 G M n_b / (a_b c^2)$ are taken into account.

Keplerian orbital elements do not constitute a canonically-conjugated set of variables. As a result, prior to applying Hamilton's equations of motion, we must first revert to cartesian Poincar\'e coordinates defined by 
\begin{equation}
h = e \cos(\varpi) \ \ \ \ \ \ k = e \sin(\varpi)
\end{equation}
This canonical variable system can be manipulated to be more succinct by representing the cartesian coordinates as the real and imaginary components of a single complex Poincar\'e variable $z$: 
\begin{equation}
\label{zdef}
z= h + \imath k = e \exp{\imath \varpi},
\end{equation}
where $\imath = \sqrt{-1}$. Accordingly, in terms of the new variables, the Hamiltonian takes on a simple compact form:
\begin{equation}
\mathcal{H}^{\rm{sec}}_{b} = \frac{A}{2} z_b z^*_b + \frac{A'}{2} (z_b z^*_c + z_c z^*_b),
\end{equation}
Applying Hamilton's equation $dz/dt = 2 \imath \partial \mathcal{H}/\partial z^*$, and augmenting the evolution to account for tidal dissipation, we obtain the secular equation of motion for planet b:
\begin{equation}
\frac{dz_{b}}{dt} = \imath A z_b+ \imath A' z_c - \frac{z_{b}}{\tau}.
\label{dzdt}
\end{equation}

Note that here, we have approximated the effect of tides as exponential decay of the eccentricity with a timescale $\tau$, given by:
\begin{equation}
\tau = \left[ \frac{21}{2} \frac{k_{2_b}}{Q_b} \frac{M}{m_b} \left(\frac{R_b}{a_b}\right)^{5} n_b \right]^{-1},
\end{equation}
where $Q$ is the tidal quality factor. This approximation is only valid if the decay of the semi-major axis is neglected \citep{1966Icar....5..375G, 1999ssd..book.....M} and the planetary spin period has synchronized with its orbital period. In the problem of interest, this turns out to be a good approximation because that at low eccentricities, the decay timescale of the semi-major axes is orders of magnitude longer than that of eccentricity, while the tidal synchronization timescale is much shorter than the eccentricity decay timescale \citep{1980A&A....92..167H}. As a result, it is likely that the system will attain equilibration before $\tau$ changes significantly \citep{2002ApJ...564.1024W}. This separation of timescales allows the decay of the semi-major axes to be treated as a subsequent, adiabatic effect \citep{2013A&A...556A..28B}.

The equation of motion for the outer planet can be constructed in a similar manner. Formally, the Hamiltonian that governs the secular motion of planet $c$, as dictated by the gravitational interactions with planet $b$ is identical to equation (\ref{Hsecb}). However, taking note of the equilibrium expression of Mardling (2007), we can further simplify the system of equations by taking advantage of the fact that fixed point solutions with $a_b/a_c \ll 1$ generally also have $e_b/e_c \ll 1$ \citep{2011ApJ...730...95B}. Consequently, in the expression for $\mathcal{H}^{\rm{sec}}_c$, we shall drop the harmonic all together, yielding:
\begin{equation}
\mathcal{H}^{\rm{sec}}_c = \frac{g_c}{2} e_c^2 = \frac{g_c}{2} z_c z^*_c.
\end{equation}

Here, $g_c$ takes the place of $A$ and is approximately given by a similar expression \citep{1999ssd..book.....M}:
\begin{equation}
g_c = \frac{3}{4} \frac{m_{b}}{M} \left(\frac{a_b}{a_c} \right)^2 n_{c}.
\end{equation}
Trivial integration of Hamilton's equation immediately yields
\begin{equation}
\label{zdef2}
z_c = \mathcal{C} \exp( \imath g_c t + \delta ),
\end{equation}
where $\mathcal{C}$ is an integration constant and $\delta$ is a (possibly complex) phase, arbitrarily defined by the reference direction of the coordinate frame. Here, for simplicity, we shall set $\delta = 0$. Recalling the definition of the complex eccentricity vector $z$ (equation \ref{zdef2}), we can readily interpret this solution as precession of the perihelion with a rate $g_c$ at constant eccentricity $e_c = \mathcal{C}$. In a direct parallel, the value of the coefficient $A$ dictates the precession rate that planet $b$ would have if the eccentricity of planet $c$ was null. Note further that $A$ simplifies to just the sum of the general relativistic and tidal precessions in the limit where $m_c \rightarrow 0$.

\begin{figure}[h!] 
   \centering
  
   \includegraphics[width=3.4in]{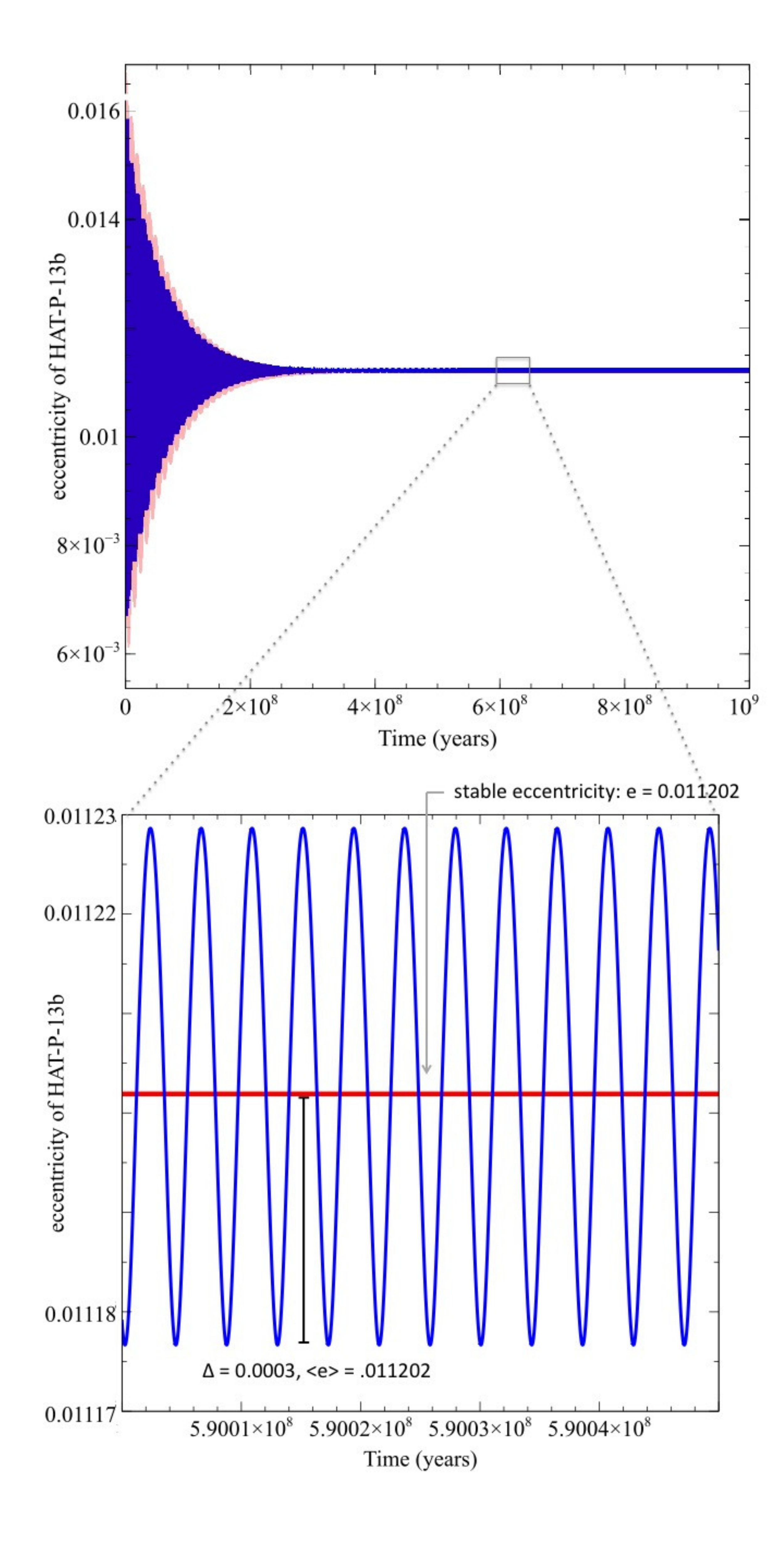} 
   \caption{After the system attains equilibrium, the variation in eccentricity behavior between the two- and three-planet cases demonstrates the difference between a fixed point (red, two-planet case) and a limit cycle (blue, three-planet case). In the case of the limit cycle, the oscillation is small enough that it is effectively a fixed point. }
   \label{selection}
\end{figure}

With a well-defined solution for $z_c$, we integrate equation (\ref{dzdt}) with respect to time to obtain a closed-form expression for the secular evolution of planet $b$: 
\begin{eqnarray}
\label{zsol}
z_b &=& \frac{A' e_c}{A - g_c + \imath/\tau} \left( \exp(\imath A t - t / \tau) - \exp(\imath g_c t) \right) \nonumber \\
 &+& \mathcal{B} \exp(\imath A t - t/\tau),
\end{eqnarray}
where $\mathcal{B}$ is an initial condition of $z_b$, defined at $t = 0$. Evidently, the dynamical evolution of the inner planet is comprised of an oscillatory component as well as an initial transient component. The characteristic timescale of the initial transient component is $\tau$. Mardling's (2007) demonstration of the fact that roughly three circularization timescales are needed for the system to attain equilibration is a manifestation of this fact.

Letting $t \gg \tau$, all the exponential terms in the solution that involve a $-t/\tau$ in the argument can be dropped. This simplifies the expression (\ref{zsol}) to
\begin{eqnarray}
\label{zbfinal}
z_b &=& - \frac{A' e_c}{A - g_c + \imath/\tau} \exp(\imath g_c t) 
\end{eqnarray}

The physical interpretation of the solution is now clear. The argument of the exponential implies that after an initial equilibration period, the inner orbit evolves into a state of co-precession with the outer orbit. Whether the periapses of the orbits end up aligned or anti-aligned depends on the relative magnitudes of $A$ and $g_c$. In particular, recalling that $A'$ is always negative, the criterion for alignment verses anti-alignment can be written down explicitly as
\begin{eqnarray}
A/g_c &>& 1 \rightarrow \varpi_b = \varpi_c \nonumber \\
A/g_c &<& 1 \rightarrow \varpi_b = \varpi_c + \pi.
\end{eqnarray}
Neglecting the tidal and general relativistic contributions to $A$, this criterion takes on a form similar to the one described by Mardling (2007).

An equivalent derivation of the same result as above can be obtained by writing down the Laplace-Lagrange secular equations of motion \citep{1999ssd..book.....M} and folding in the dissipative term into the diagonal interaction coefficients $A$ and $g$ \citep{2002ApJ...564.1024W}. In this formulation, the solution is given by a superposition of two linear eigenmodes and the approach to apsidal alignment is represented by the (relative) decay of one of the eigenmodes.

Note that strictly speaking, the solution does not imply exactly aligned or anti-aligned orbits. The complex component of the denominator implies that the orbits are actually misaligned by $\Delta \varpi \simeq (\tau (A-g))^{-1}$ \citep{2007PhDT.........3Z}. However, provided that tidal circularization is generally much slower than the secular interaction timescale of the system, $\Delta \varpi$ is typically small enough to be neglected in practice. 

In the above formulation, we have explicitly neglected tidal decay of the semi-major axes invoking the separation of timescales as justification. Now that we have expressed a long-term solution for the eccentricity of the inner planet, we can readily account for the neglected effect. Once again taking advantage of the fact that tidal evolution is slow in comparison to secular evolution, we assert that the area enclosed by a single secular cycle in phase space is an adiabatic invariant and is therefore conserved. Noting that such area is essentially null when the system resides at a fixed point, this conservation implies that tidal decay of the semi-major axis does not perturb the system away from equilibrium.

Because the eccentricity is now quasi-steady, the rate of change of the semi-major axes due to tides raised on the planet reads:
\begin{eqnarray}
\frac{da}{dt} = -2 z_b z_b^* \frac{a}{\tau}
\end{eqnarray}
Provided a sufficient amount of time, this decay may lead to considerable changes in the relative magnitudes of $A$ and $g_c$. In particular, cases where the frequencies become resonant may be envisioned. In such situations, passage through the linear resonance gives rise to a temporary excitation of eccentricity, thanks to the singularity in expression (\ref{zbfinal}) (Mardling 2007). Note that the tides raised on the (slowly rotating) star will also contribute to shrinking the semi-major axes. The additional dissipation can be included into any practical calculation but will not change the behavior of the system qualitatively from the picture described here.

\section{Dissipative Evolution of a Giant Planet, Perturbed by a System of Distant, Massive Objects}

In the previous section, we set the stage for the following calculation by considering a well-studied example of the dissipative secular three-body problem. In this section we shall consider an extension of the previous calculation and explore the behavior of the fixed point under perturbations from an additional planetary companion. In other words, we would now like to establish the criteria for 4-body planetary orbital architectures, where the deviations away from the secular fixed point of the inner planetary pair are negligible.

In direct analogy with the previous section, accounting for interactions with planet $d$, the Hamiltonian that governs the secular evolution of the inner most planet reads
\begin{equation}
\label{Hsecb2}
\mathcal{H}^{\rm{sec}}_b = \frac{\tilde{A}}{2} z_b z_b^* + \frac{A'}{2} (z_b z^*_c + z_c z^*_b) + \frac{A''}{2} (z_b z^*_d + z_c z^*_d),
\end{equation}
where
\begin{eqnarray}
\tilde{A} &=& \left(\frac{d\varpi}{dt}\right)_{\rm{tidal}} + \left(\frac{d\varpi}{dt}\right)_{\rm{GR}} + \frac{3}{4} \frac{m_{c}}{M} \left(\frac{a_b}{a_c} \right)^3 n_{b} \nonumber \\
&+& \frac{3}{4} \frac{m_{d}}{M} \left(\frac{a_b}{a_d} \right)^3 n_{b},  \nonumber \\
A' &=&-\frac{15}{16}  \frac{m_{c}}{M} \left(\frac{a_b}{a_c} \right)^4 n_{b}, \nonumber \\
A'' &=&-\frac{15}{16}  \frac{m_{d}}{M} \left(\frac{a_b}{a_d} \right)^4 n_{b}.
\end{eqnarray}
As before, in our description of the orbital evolutions of the two outer planets, we shall approximate the secular effect of the inner-most planet as that arising from a circular ring of mass. However, the secular interactions of the outer two planets will be treated in a more self-consistent way. Specifically, we shall derive the evolution of the state vectors of the outer planets directly from the Laplace-Lagrange secular theory \citep{1996CeMDA..64..115L, 1999ssd..book.....M, 2011ApJ...739...31L}.

As already mentioned above, mathematically the Laplace-Lagrange theory constitutes a regular eigenvalue problem. Consequently, the secular solution for the two outer planets can be written generally as 
\begin{eqnarray} 
\label{LLsol}
z_{c} &=& \beta_{1,1} \exp{(\imath g_{1}t+\delta_{1})} +  \beta_{1,2} \exp{(\imath g_{2}t+\delta_{2})} \nonumber \\
z_{d} &=& \beta_{2,1} \exp{(\imath g_{1}t+\delta_{1})} +  \beta_{2,2} \exp{(\imath g_{2}t+\delta_{2})}
\end{eqnarray}
where $g$'s, and $\beta$'s are the eignenvalues and scaled eignvectors and of a matrix comprised by interaction coefficients $A$, while $\delta$'s are the phases dictated by initial conditions \citep{1999ssd..book.....M}.  

Note that because no direct or implicit dissipation is applied to the outer planets, the eignenvalues $g_1$ and $g_2$ are real. This is in contrast to the usual treatment of the problem where all three planets are considered explicitly and tidal damping (that acts only on the innermost planet) introduces imaginary components into all eigenvalues. If the associated decay timescales of all but one eigenmodes are short compared to the age of a given system, such a system evolves to a global fixed point, characterized by parallel periapses \citep{2011ApJ...730...95B}. However, here we wish to focus on a converse scenario (i.e. a case of well-separated outer orbits) where the decay of only a single eigenmode is short and the outer orbits do not attain (anti-)alignment as a result of dissipation.

Explicitly, the equation of motion for the innermost planet reads:
\begin{eqnarray} 
\frac{d z_b}{dt} &=& \imath (\tilde{A}+\frac{\imath}{\tau}) z_b + \imath (A' \beta_{1,1} + A'' \beta_{2,1}) \exp (\imath g_1 t + \delta_1) \nonumber \\
&+& \imath (A' \beta_{1,2} + A'' \beta_{2,2}) \exp (\imath g_2 t + \delta_1).
\end{eqnarray}
The equation admits the solution
\begin{eqnarray}
\label{twomodes}
z_b &=& \frac{ A' \beta_{1,1} + A'' \beta_{2,1}  }{\tilde{A} - g_1 + \imath/\tau } \big( \exp(\imath \tilde{A}t +\imath \delta_1 - t/\tau)  \nonumber \\
&-& \exp(\imath g_1 t + \imath  \delta_1)   \big) + \frac{A' \beta_{1,2} + A'' \beta_{2,2}}{\tilde{A} - g_2 - \imath/\tau } \nonumber \\
&\times&( \exp(\imath \tilde{A} t + \imath \delta_2 + t/\tau) - \exp(\imath g_2 t + \imath  \delta_2) ).
\end{eqnarray}

Dropping the transient terms, the solution remains a linear super-position of two modes. Although the particularities of the solution depend on the system in consideration, typically one of the modes will dominate over the other. Because the interaction coefficients $A'$ and $A''$ depend strongly on the semi-major axis ratio, $A' \gg A''$ in well-separated systems (unless $m_d \gg m_c$). More intuitively, dropping the $A''$ terms from the Hamiltonian (\ref{Hsecb2}) but using the solution (\ref{LLsol}) for $z_c$ is equivalent to neglecting the gravitational interactions between planets $b$ and $d$.

To obtain a better handle on the dynamical portrait of the inner planet's motion, it is useful to transform to a coordinate system which drifts along with one of the modes. The advantage of doing so is the removal of time-dependance from one of the exponentials in equation (\ref{twomodes}). Taking the inner-pair gravitational interactions to be dominant, it is sensible to transform to a variable system defined by 
\begin{equation}
\bar{z}_b = z_b \exp(- \imath \varpi_1) =  e_b \exp(\imath \varpi_b - \imath g_1 t - \imath \delta_1).
\end{equation}
Upon doing so and dropping the small $\imath/\tau$ terms in the denominators of equation (\ref{twomodes}), the $t \gg \tau$ secular solution for planet $b$ reads:
\begin{eqnarray}
\label{barzb}
 \bar{z_{b}}  &\equiv& \left< e \right> + \Delta \exp (\imath \varphi t + \imath \gamma) \nonumber \\
  &=& \frac{ A' \beta_{1,1} + A'' \beta_{2,1}  }{g_1 - \tilde{A}} + \frac{A' \beta_{1,2} + A'' \beta_{2,2}}{g_2 - \tilde{A} } \nonumber \\
 &\times& \exp(\imath (g_2 - g_1) t + \imath (\delta_2 - \delta_1) ) \nonumber \\
\end{eqnarray}

The above solution describes an inner planet with a constant mean eccentricity $\left< e \right>$, precessing at the rate of mode 1 on average. Note that information about the planetary Love number is still embedded in the solution through $\tilde{A}$. However, the state vector of the planet now executes a limit cycle of width $\Delta$ instead of staying fixed as in the case of the three-body problem. It would seem that the associated variation spoils the goal of indirectly quantifying $k_{2}$ by measuring the orbital eccentricity as the observable quantify evolves in time, even after orbital equilibration.

This complication is remedied by the fact that the quasi-stationary nature of the inner planet's dynamical state is maintained to the extent that $\Delta \ll \left< e \right>$. Furthermore, in the limit of small deviations from the fixed point, the maximal apsidal deviation away from the periapsis described by mode 1 is simply given by $|\varpi_b - \varpi_1| \simeq \Delta / \left< e \right>$. In other words, the above expression dictates a simple criterion for an effectively fixed nature of the inner planetary pair's dynamics: 
\begin{eqnarray}
e_b^{\rm{fixed}} = \left< e \right>  \ \ \ \ \ \rm{if} \ \ \Delta \ll \left< e \right>.
\end{eqnarray}
Indeed, the deviation away from the fixed point may be negligible compared to the observational uncertainties. In such cases, the intrinsic errors in the constraints on the planetary interior structure will be not be dominated by dynamical effects and the calculation can proceed as in the three-body case \citep{2009ApJ...704L..49B, 2012A&A...538A.146K} to a satisfactory accuracy.

An example of a system of this sort is presented in Figure \ref{selection}. While the orbital parameters of the inner planet pair are adopted from the observed parameters of the HAT-P-13 system (Table \ref{params}), the third planet is taken to have an eccentricity of $e_{d} = 0.3$ and a semi-major axis of $a_{d} = 10 AU$. 

An intuitive interpretation of the above results can be presented in terms of adiabatic theory (Henrard 1982). Specifically, if the timescale for (any) external perturbation to the innermost planetary pair greatly exceeds the timescale for secular exchange of angular momentum among the two inner-most planets (a quantity closely related to $\tilde{A}$), the action (which here is referred to as the adiabatic invariant)
\begin{equation}
J = \oint \sqrt{G M_{\star} a_b} (1 - \sqrt{1 - e_b^2}) \ d (\Delta \varpi)
\end{equation}
will remain a quasi-conserved quantity. 
Physically, $J$ represents the area occupied by the trajectory in phase-space. If the orbit in question resides on a fixed point, the associated adiabatic invariant is identically null: $J_{\rm{fixed}} = 0$. $J$'s value is indicative of how closely the system adheres to an apsidally aligned state. 
It is noteworthy the adiabatic invariant, as defined above, is actually the leading order approximation to the "real" adiabatic invariant, which can be calculated using the Lie perturbation series approach (e.g. Lightenberg \& Lieberman 1992, Henrard, 1974). Furthermore, while some authors choose to retain a strict definition of adiabatic forcing which corresponds exclusively to modulation of the amplitude of the harmonic in the Hamiltonian, here we adopt a more crude, but more widely used definition of any slow perturbation, which in turn allows for the modulation of the locations of equilibrium points in phase space (Goldreich \& Peale 1966, Henrard 1982, Peale 1986). 

In the context of this interpretation, an equivalent criterion for the quasi-stationary nature of the solution may be formulated:
\begin{eqnarray}
e_b^{\rm{fixed}} = \left< e \right>  \ \ \ \ \ \mathrm{if} \ \ \tilde{A} \gg (g_1,g_2).
\end{eqnarray}

\begin{figure}[h!] 
   \centering
   \includegraphics[width=3.4in]{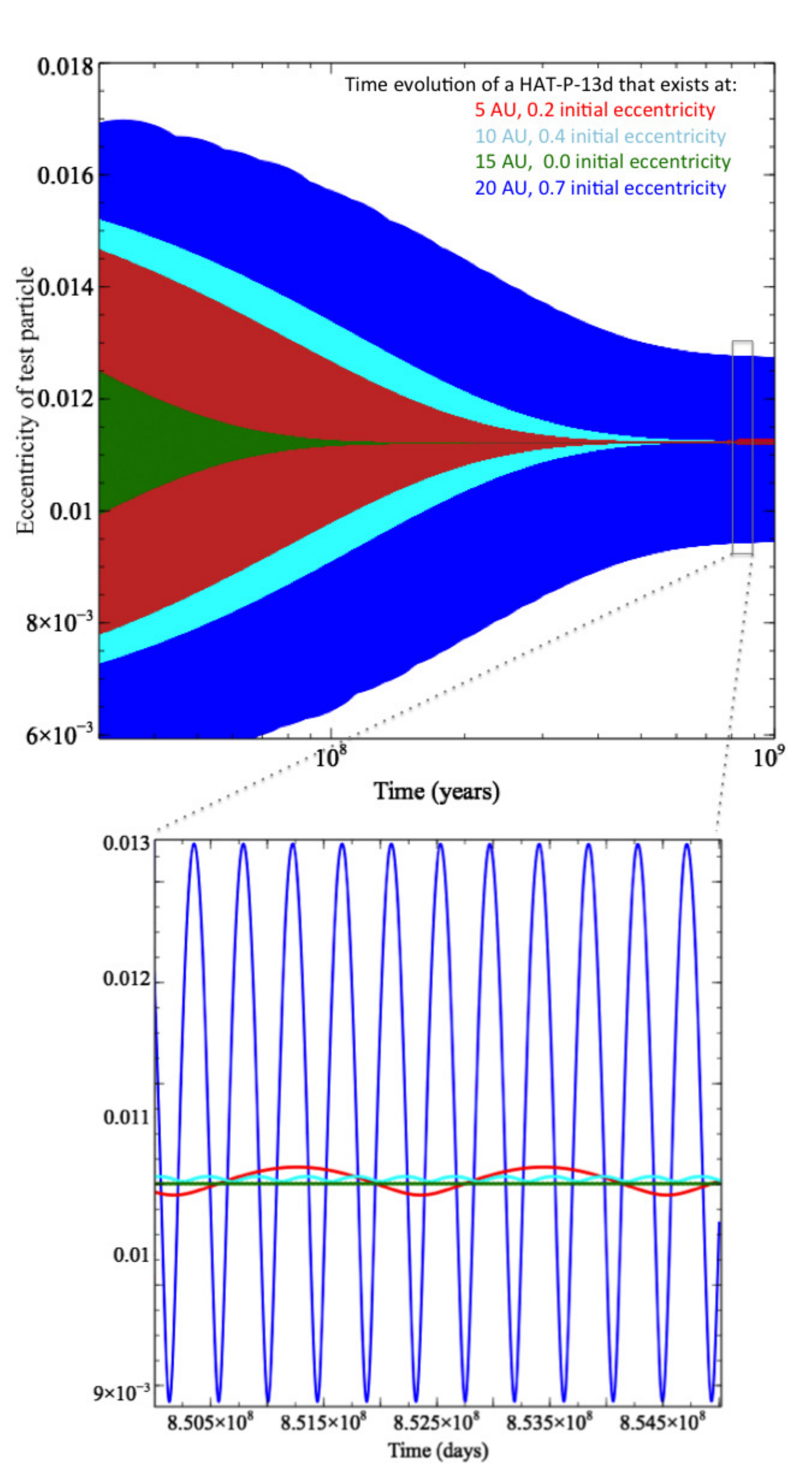} 
   \caption{The amplitude of the oscillation in eccentricity, relative to the eccentricity of the innermost planet itself, must be small for the planet to be considered at an effective fixed point. The semi-major axis and eccentricity of the third planet in the system determines whether the innermost planet resides at a fixed point or not. Examples of systems at effective fixed points are shown in the figure above, in green, cyan, and red. In these cases, the amplitude of oscillation in eccentricity for the innermost planet is small relative to the value of eccentricity itself. In contrast, an example of a system not at an effective fixed point is given in blue, where the eccentric orbit of the third planet increases the oscillation amplitude of the innermost body. The eccentricities of the outermost planet is exaggerated in the most distant (higher semi-major axis) case for illustrative purposes, regardless of the stability of that configuration.}
   \label{fig3}
\end{figure}

\section{An Application to the HAT-P-13 System}
The story of the perfect test case for this methodology begins with the Hungarian Automated Telescope Network (HATNet) survey \citep{2002PASP..114..974B}. This system of six automated 11cm telescopes operates with a goal of finding transiting exoplanets. The transiting inner planet (HAT-P-13b) was found by the HATNet system, and follow-up radial velocity work at Keck by \cite{2009ApJ...707..446B} confirmed a second, non-transiting planet (HAT-P-13c) existed in the system. 

While HAT-P-13 was thought to be a system of two planets, the observed apsidal alignment between planets $b$ and $c$ signaled orbital equilibration. Upon inspection, the system appeared to pose an excellent example of the three-body system described in section 3. As such, \cite{2009ApJ...704L..49B} used an octopole-order (in semi-major axis ratio) expansion of the secular Hamiltonian, formulated by \cite{2007MNRAS.382.1768M}, to place an upper limit of $\sim 120 M_{\oplus}$ on the planetary core mass.  

The story of HAT-P-13 was not yet finished, however; long-term radial velocity monitoring of the system points to the existence of a third giant planet (or possibly brown dwarf) in the system. Due to the relatively long-period, faint signal present in radial velocity data, the orbit and mass of this third perturber are largely unconstrained. However, the data yields an order of magnitude constraint in the relationship between the mass and semi-major axis of the (assumed coplanar) planet \citep{2010ApJ...718..575W}:
\begin{equation}
\Bigg( \frac{M_{d} }{M_{\mathrm{Jup}}}\Bigg) \Bigg( \frac{a_{d}}{10 \textup{AU}} \Bigg)^{-2} \simeq \textup{9.8}
\end{equation}
where $M_{d}$ refers to the mass of HAT-P-13d and $a_{d}$ refers to its semi-major axis. The orbital and physical parameters of the planetary system are summarized in Table \ref{tab2}. 

This newly acquired data renders HAT-P-13 an ideal example of a 4-body system, as considered in section 4. As such, the existence of a third planet begs the previously posed question: are the interior structure determination methods used by \cite{2009ApJ...704L..49B} still valid for this system, given that there is an additional perturbing body? As already discussed in the previous section, the answer to this question is given by the width of the limit cycle, $\Delta$, relative to $\left< e \right>$. A small value of $\Delta / \left< e \right>$ could mean that a quasi-stationary solution for the inner orbit continues to be a possibility. As a result, this section will focus on the delineation of the orbital solutions for planet $d$ that allow for the interior structure calculation to remain well-founded.

To determine the regime in which such a solution could still exist, we consider a sequence of planar system architectures where the outermost planet's orbital parameters are chosen randomly with the exception of the semi-major axis and mass, that are subject to radial velocity constraints (see Table \ref{tab2}).

\begin{figure}[h!] 
   \centering
   \includegraphics[width=3.4in]{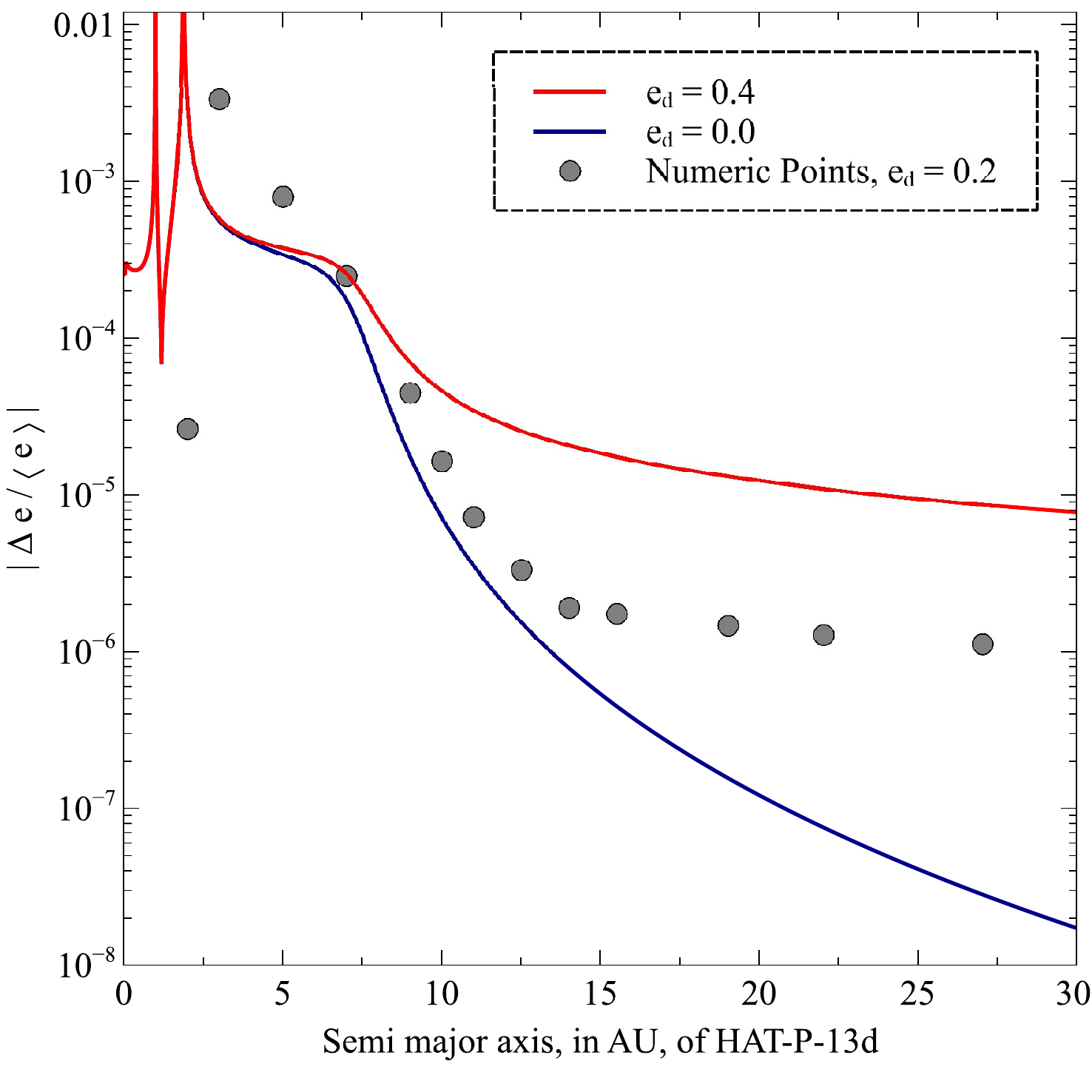} 
   \caption{The analytic solutions for two representative eccentricities (the upper and lower limits of stability, as illustrated in figure \ref{stability}) both yield a narrow width of limit cycle across all semi-major axis of the third planet. Numerical points for a generally stable eccentricity confirm this solution to order of magnitude. The third planet's orbit becomes unstable when its semi-major axis and eccentricity are such that orbit crossing occurs with the inner planets. } 
   \label{f2}
\end{figure}

Considering a semi-major axis range of 2AU - 30AU, we computed the quantity $\Delta/\left<e \right>$ as given by equation (\ref{barzb}). Interestingly, we found the results to be relatively insensitive to the outermost orbit's eccentricity and representative analytic solutions with $e_{d} = 0.0$ and $e_{d} = 0.4$ are given in figure (\ref{f2}). 

In the case that $a_{d}$ is chosen such that the third planet's orbit resides near those of the inner two planets, there is significant instability.

An example of the the regime in which quasi-stationary solutions (that with a small value of $\Delta/\left<e \right>$ ) exist is shown in Figure \ref{f2}. Additionally, Figure (\ref{f2}) includes a self-consistent numerical solution obtained with an N-body code. In the latter approach, the equations of motion were solved using the Bulirsch-Stoer method \citep{1986nras.book.....P} while the tidal formalism of \cite{2002ApJ...573..829M} was used to account for the dissipative effects. General relativistic effects were taken into account by incorporation of an ad-hoc potential term as shown in \cite{2001ApJ...554.1141L}.  

Although the analytical results presented in Figure (\ref{f2}) are qualitatively consistent, it is important to keep in mind that through out the derivations presented above, only the leading terms in the expansion of the secular Hamiltonian were retained. That is to say that on a quantitative level, the theory should only be viewed as a leading order approximation. In order to confirm our estimates, we performed a series of numerical experiments where a chosen orbital configuration was integrated numerically with a dissipative N-body code described above over many circularization timescales and allowed to attain orbital relaxation. This is shown in Figure (\ref{fig3}). Note the consistency in the limit cycles for planets with stable orbits; in Figure (\ref{fig3}), the orbit with a semi-major axis of 5 AU experiences a slightly larger value of  $\Delta/\left<e \right>$, consistent with the observed effect in Figure (\ref{f2}).

As a final exercise, we considered the dynamical stability of the putative 3-planet HAT-P-13 system with the aim to constrain the orbital state of the outermost planet. This was done by a performing a series of conservative N-body simulations using the Mercury6 software package (Chambers 1999) over a few secular periods. The hybrid algorithm was used throughout. As before, the starting conditions for the third planet were chosen randomly. 

The orbital architecture of HAT-P-13 is such that orbital instabilities that arise from arbitrarily chosen initial conditions essentially always stem from interactions of the outer pair of orbits. Naively, it is tempting to neglect the innermost planet altogether and simply integrate the 3-body system. This is however unwarranted, since the presence of the inner planet may have an appreciable secular effect on the outer orbits. Analogous to the case of general relativistic precession and the dynamical stability of Mercury, the induced apsidal precession may even be stabilizing in some circumstances. At the same time, the innermost planet's short period complicates numeric simulations since the time step needed to correctly resolve the innermost orbit is about two orders of magnitude shorter than that needed for the orbit of planet $c$.

Taking advantage of the large orbital separation of the inner pair of orbits, we reconciled this issue by noting that the orbit-averaged gravitational effect of HAT-P-13b is analogous to an inertially equivalent rotational bulge on the star. Consequently, our simulations comprised planets $c$ and $d$ in orbit of an oblate primary. The added stellar oblateness (in the form of $J_{2}$) was calculated such that its precessional effect mimics that produced by the leading order term in the secular Hamiltonian:
\begin{eqnarray}
g_c=\frac{3}{2}J_{2}\left(\frac{R_{*}}{a} \right)^{2}-\frac{9}{8}J_{2}^{2}\left(\frac{R_{*}}{a}\right)^{4}
\end{eqnarray}

\begin{figure}[h!] 
   \centering
   \includegraphics[width=3.4in]{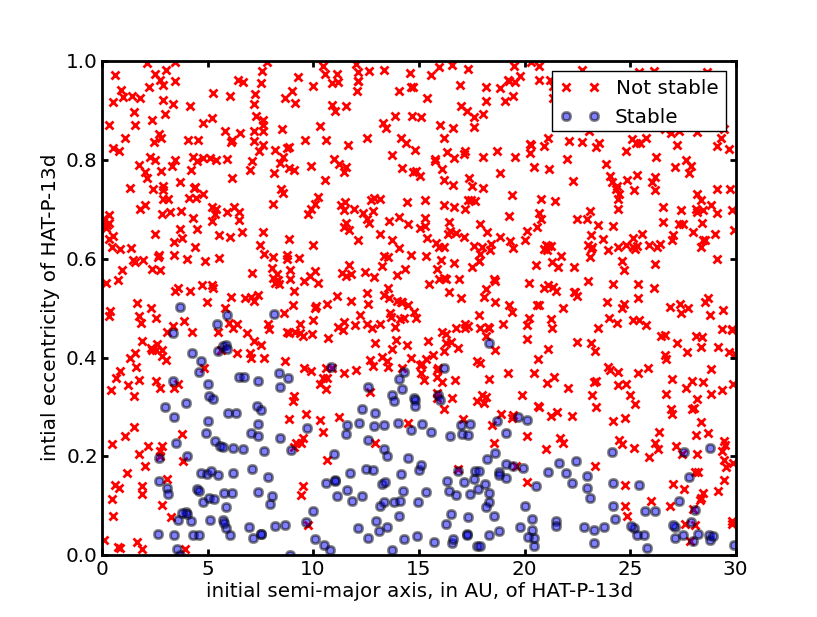} 
   \caption{A Monte Carlo simulation helped identify the stable and unstable regions in $a - e$ space for the third planet. Unstable points, in red, indicate that a planet in the system is ejected or crashed into the star when a third planet with the given semi-major axis and eccentricity was added to the system. Stable points, in blue, indicate that the system is stable for two secular periods. Given that the third planet must exist in this stable region, and thus likely behaves adiabatically with respect to the inner two-planet system, it is likely that this outer planet does not prevent an interior structure estimation from being possible. }
   \label{stability}
\end{figure}

The stability map of the HAT-P-13 system is shown in Figure (\ref{stability}) where the blue and red points denote stable and unstable initial conditions respectively. The Figure demonstrates a clear pattern: the eccentricity of the outer-most planet cannot exceed $\sim 0.4$ and the allowed ellipticity of the outer orbit decreases rapidly for $a_d > 20$AU. That said, we are aware of the fact that our investigation of dynamical stability is far from exhaustive, since orbital configurations corresponding to mean motion resonances can yield stable orbits at high eccentricity (Correia et al 2009) and we are not treating the initial conditions with sufficient care to identify such states. As an extension of the parameter survey, we also considered mutual inclination between the planets. However, the stability map in the $e - a$ plane was not modified significantly. It is interesting to note that the combined interpretation of Figures (\ref{f2}) and (\ref{stability}) is that system architectures that are well below the stability boundary also behave adiabatically with respect to the secular dynamics of the inner planetary pair. Consequently, we conclude that HAT-P-13b remains an excellent candidate for the estimation of the interior structure of an extra-solar planet.

\section{Conclusion}

In this study we have considered the dissipative dynamical evolution of secular multi-planet systems and investigated the viability of indirect measurements of transiting planetary structure. We began our discussion by formulating the utility of the planetary Love number as a means to constraining the interior structure, using a closed-form polytropic model of the planet. In particular, we argued that the Love number, when observable, can provide a clue towards the presence or absence of a core, whereas the planetary radius (in the $\sim 1 - 10 M_{\rm{Jup}}$ range) is almost solely dictated by overall metallicity of the object. 

Subsequently, we used simple secular perturbation theory to analytically reproduce the now well-known process of orbital equilibration and approach to an apsidally co-linear fixed point. By doing so, we were able to write down a simple equilibrium equation that explicitly demonstrates the dependence of the eccentricity on the planetary Love number, illuminating the indirect observational avenue to constraining extrasolar planets' interiors, as in \cite{2009ApJ...704L..49B}. We then expanded our discussion to incorporate perturbations from an additional body and showed that in some similarity to the discussion of \cite{2010MNRAS.407.1048M}, the fixed point is replaced by a limit-cycle in the ($e,\Delta \varpi$) plane. However, if the characteristic timescale of the external perturbation is taken to be long compared to the characteristic interaction timescale of the inner planetary pair, the perturbation acts in an adiabatic fashion, yielding a limit-cycle with a negligible width.

As an application of the formulated theory, we considered the dynamical evolution of the HAT-P-13 system, which has become the canonical example used for the estimation of extrasolar interior structure. Exploring a range of orbital architectures loosely constrained by the radial velocity data, we showed that except for a narrow portion of parameter space, the inferred presence of an additional massive companion does not spoil the calculation of the innermost planet's interior structure. Additionally, using dynamical stability constraints, we placed weak restrictions on the orbital state of the uncharacterized planet, arguing that its orbital eccentricity must be mild.

Although the discussion in this paper describes the secular evolution of a planetary pair perturbed by a single additional companion, the employed method can be easily extended to numerous perturbers. In such a case, the perturbing system of N bodies (also assumed to be dominated by secular interactions) will be governed by N eigenmodes. Accordingly, the fixed-point dynamics will also be modulated by N terms and the resulting limit-cycle will have a complicated shape. However, provided that all the perturbations are slow, the cumulative width of the limit cycle can remain inconsequential.

Within the realm of this work, the aim was to explore and demonstrate the limit-cycle behavior of the eccentricity dynamics in the adiabatic regime. Similar arguments apply to mutually inclined systems. Such a problem was recently investigated by \cite{2012Natur.491..418B}. Unfortunately, simultaneous analytical treatment of eccentricity and inclination dynamics is made difficult by the quartic coupling terms in the secular Hamiltonian \citep{1999ssd..book.....M}. However, the adiabatic principle affirms that a coplanar planetary pair, subject to slow external perturbation will maintain its coplanarity. It is likely that orbital equilibration and the associated possibility of inferring the planetary Love number will not be affected by long-term perturbations of any kind. 

As a concluding remark, it seems worthwhile to comment on the prospects of the determination of $k_2$ given the current knowledge of the orbital distribution of extrasolar planets. Even prior to the release of the vast dataset obtained by the \textit{Kepler} spacecraft \citep{2012AAS...22030601B}, conventional radial velocity and transit surveys showed a relatively sharp distinction between systems hosting hot Jupiters and systems hosting hot sub-Neptune mass planets. While hot Jupiters are rarely accompanied by planets whose orbits reside within $\sim 1$AU \citep{2010arXiv1006.3727R}, compact multi-planet systems systems of Super-Earths are quite common \citep{{2010SPIE.7735E..33L}}. 

The requirements for the measurement of $k_2$ in the hot Jupiter case are quite clear. Namely, the maintenance of the transiting planet's orbital eccentricity in face of tidal dissipation by the perturbing planet requires it to lie on an orbit with much more angular momentum i.e. $(a_b/a_c) \ll 1$. Additionally, an enhanced orbital eccentricity of the inner planet is favorable from an observational perspective. Recalling that $(e_b/e_c)_{\rm{fixed}}\propto (a_b/a_c)$, this constraint renders a highly eccentric outer planet favorable. Finally, as argued by \cite{2010MNRAS.407.1048M}, coplanarity of the inner-most orbital pair is a must. 

The case of systems of hot low-mass planets is considerably more unfortunate. Because the angular momentum budget of the entire system does not exceed that of the inner most planet by an overwhelming amount, all secular modes tend to decay away rapidly, leading to near-circular orbits. In turn, this process guides the systems towards a state dominated by the resonant normal form \citep{2013A&A...556A..28B,2012ApJ...756L..11L,2012A&A...546A..71D}. Owing to the near-circularity of the orbits, a distinctive feature of such a dynamical state is the rapid retrograde recession of the longitudes of perihelia of the planets. Indeed, such a recession completely dominates over tidal or any other non-Newtonian source of precession. Even in systems that avoid rapid circularization (e.g. 61 Virginis, \cite{2010ApJ...708.1366V}), the tidal precession term is bound to be relatively small, due to its extreme sensitivity on the planetary radii. In other words, the effective signal-to-noise ratio of the tidal effects to the overall dynamical state is essentially negligible in hot low-mass systems.

The discussion suggests, that planetary systems with orbital architectures mirroring that of HAT-P-13 appear to be ideal candidates for indirect estimation of extrasolar interior structures. Consequently, continued radial-velocity monitoring of transiting hot Jupiters in search for distant massive companions is essential for the acquisition of a theoretical understanding of extrasolar planetary interiors. 

\section{Acknowledgements}

We thank John Johnson for his careful review of the manuscript and helpful suggestions. We would additionally like to thank Greg Laughlin for useful conversations. K. Batygin acknowledges the generous support of the ITC Prize Postdoctoral Fellowship. We also thank the anonymous referee for their insightful report.

\clearpage
\section{Appendix}
\begin{table}[h]
\caption{HAT-P-13 Observed System Properties \citep{2009ApJ...707..446B}} 
\centering 
\begin{tabular}{c c c c c c} 
\multicolumn{6}{c}{Solar Properties} 
\\ \hline 
&Star & Radius  & Mass  & Magnitude  &  \\ [0.1ex] 
\hline 
&HAT-P-13  & 1.56ÊÊÊÊÊÊÊÊÊÊÊÊÊÊÊÊ & 1.22ÊÊÊÊÊÊÊÊÊÊÊÊÊÊ & 10.62ÊÊÊÊÊÊÊÊÊÊÊÊÊÊ  \\[1ex]    
\hline
\multicolumn{6}{c}{Planetary Properties} \\[1ex]
\hline 
Planet & Radius  & Mass  & Semi-major axis (AU) & Eccentricity  \\ [0.1ex] 
\hline 
HAT-P-13b & 0.85ÊÊÊÊÊÊÊÊÊÊÊÊÊÊÊÊ & 0.85ÊÊÊÊÊÊÊÊÊÊÊÊÊÊ & 0.0426ÊÊÊÊÊÊÊÊÊÊÊÊÊÊ & 0.0142ÊÊÊÊÊÊ &\\ 
HAT-P-13c & -ÊÊÊÊÊÊÊÊÊÊÊÊÊÊÊÊÊÊÊ & 14.5ÊÊÊÊÊÊÊÊÊÊÊÊÊÊ & 1.186ÊÊÊÊÊÊÊÊÊÊÊÊÊÊÊ & 0.666ÊÊÊÊÊÊÊ  \\
HAT-P-13d & -ÊÊÊÊÊÊÊÊÊÊÊÊÊÊÊÊÊÊÊ & $M_{d} = 9.8 M_{Jup}(\frac{a_{d}}{10 AU})^{2}Ê$
ÊÊÊÊÊÊÊÊÊÊÊÊÊÊ & $a_{d} = (\frac{(10 AU)^{2}}{9.8} \frac{M_{d}}{M_{Jup}})^{\frac{1}{2}}$ÊÊÊÊÊÊÊÊÊÊÊ& - 	 Ê \\[0.1ex]    
\hline
\end{tabular}
\label{tab2}
\label{params} 

\end{table}

\end{document}